\documentclass[preprint,onecolumn,nofootinbib]{revtex4-1}
\usepackage[colorlinks=true,linkcolor=blue,urlcolor=blue,filecolor=black,citecolor=red,pdfstartview=FitV,pdftitle={},pdfsubject={},pdfkeywords={},pdfpagemode=None,bookmarksopen=true]{hyperref}
\usepackage{graphicx}
\usepackage{amsmath}
\usepackage{amssymb,ulem}
\usepackage{color}%
\usepackage{tikz}
\usepackage{dcolumn}
\usepackage{bbold}
\usepackage{xcolor}

\begin{document}
\title{Dynamical spontaneous scalarization in Einstein-Maxwell-scalar theory}
\author{Wei Xiong $^{1}$}
\email{phyxw@stu2019.jnu.edu.cn}
\author{Peng Liu $^{1}$}
\email{phylp@email.jnu.edu.cn}
\author{Chao Niu $^{1}$}
\email{niuchaophy@gmail.com}\thanks{corresponding author}
\author{Cheng-Yong Zhang $^{1}$}
\email{zhangcy@email.jnu.edu.cn}
\author{Bin Wang $^{2,3}$}
\email{wang\_b@sjtu.edu.cn}

\affiliation{
  $^1$ Department of Physics and Siyuan Laboratory, Jinan University, Guangzhou 510632, China
}
\affiliation{
  $^2$ Center for Gravitation and Cosmology, College of Physical
  Science and Technology, Yangzhou University, Yangzhou 225009, China
}
\affiliation{
  $^3$ School of Aeronautics and Astronautics, Shanghai Jiao Tong
  University, Shanghai, China
}

\begin{abstract}
  We study the linear instability and the nonlinear dynamical evolution of the Reissner-Nordstr\"om  (RN) black hole in the Einstein-Maxwell-scalar theory in asymptotic flat spacetime. We focus on the coupling function $f(\phi)=e^{-b\phi^2}$ which allows both the scalar-free RN solution and scalarized black hole solution. We first present the evolution of system parameters during dynamic scalarization. For parameter regions where spontaneous scalarization occurs, we find that the evolution of the scalar field at the horizon is dominated by the fundamental unstable mode from linear analysis at early times. At late times, the nonlinear evolution can be viewed as the perturbation of scalarized black holes.
\end{abstract}
\maketitle
\tableofcontents

\section{Introduction}
\label{sec:introduction}

General relativity (GR) has achieved great success both in theory and experiment, and there is a no-hair theorem that restricts that black holes can be completely determined by three parameters: their mass, charge, and angular momentum. However, it must describe the current universe through unknown physics such as dark matter and dark energy. The non-renormalizable singularity introduced in GR is also one of the difficulties. In recent years there are many theories beyond GR, some of which circumvent the
no-hair theorem. The dilatonic or colored hairy black holes were observed in the Einstein-dilaton-Gauss-Bonnet theory  and the higher dimensional or rotating hairy black hole solutions were found \cite{Torii1997,Kanti1997,Kleihaus2011,Kleihaus2015,Pani2011,Herdeiro2014,Ayzenberg2014,Sotiriou2014}. Or even GR with a certain matter sources (Yang-Mills field \cite{HHVolkov1989,HHBizon1990,HHGreene1993,HHMaeda1994} , Skyrme field \cite{HHLuckock1986,HHDroz1991}, conformally-coupled scalar field  \cite{HHBekenstein1975})
or non-minimally coupled scalar field allows the hairy black hole solutions. We are interested in the hairy black hole solution of the latter situation produced by the dynamic mechanism called spontaneous scalarization \cite{Damour1993,Damour1996,Harada1997,Cardoso2013,Zhang2014}.

The spontaneous scalarization is typically caused by a non-minimal coupling between a real scalar field and some source term, which could be the geometric invariant sources (the Ricci scalar \cite{Herdeiro2019}, Guass-Bonnet \cite{Doneva1711,Silva1711,Antoniou1711,Cunha1904,Dima:2020yac,Herdeiro2009,Berti2009}, Chern-Simon invariant \cite{Brihaye2018}) or the matter invariant sources (Maxwell invariant \cite{Herdeiro:2018wub,Guo2021}). This non-minimal coupling introduces the tachyonic mass for the scalar in a certain parameter range. Given a sufficiently small scalar perturbation, the tachyonic instability can be triggered and provide exponential growth of scalar fields. A black hole with scalar hair is formed in the final state. The dynamical mechanisms of spontaneous scalarization has been widely investigated in extended Scalar-Tensor-Gauss-Bonnet (eSTGB) theory \cite{Ripley2019,Ripley2019b,Ripley2020,Ripley2020b,Ripley2021,Silva2020,Kuan2021}. 

The EMS theory with non-minimal coupling $e^{-b\phi^{2}}$ between scalar and Maxwell field is of interest to us. This theory can return to purely Einstein-Maxwell theory when we set $\phi \rightarrow 0$. The stability of the EMS theory has been studied in \cite{Herdeiro:2018wub,Fernandes2019} and the existence of spontaneous scalarization has been confirmed in this theory. The nonlinear evolutions of scalarization induced by Maxwell field have been done in asymptotic AdS spacetime and also in asymptotic flat spacetime with different coupling functions \cite{Hirschmann2017,Zhang2021,Zhang:2021ybj,Fernandes2019b,Zhang2021b,Zhang:2021nnn}. In \cite{Herdeiro:2018wub,Fernandes2019}, Herdeiro and Fernandes also evolve the unstable RN black hole in the EMS system. However, they mainly focus on the comparison between the dynamical endpoint of the evolution and the static solutions, and the final state of evolution under the non-spherical perturbation.

We here first present the nonlinear dynamic evolution of the spontaneous scalarization of the EMS model in the asymptotic flat spacetime in detail. We start from the linear level to investigate the tachyonic instability by the continue fractions method. Then we trigger the spontaneous scalarization by introducing a small initial value of the scalar field on the background of the RN black hole for the nonlinear evolution and investigate the evolution of various parameters during the dynamic scalarization. We find that the second evolution stage, when the scalar field grows exponentially, is dominated by the fundamental tachyonic unstable mode.

This paper is organized as follows. In section \ref{section2} we introduce the Einstein-Maxwell-scalar theory and the equations of motion for evolution. Next section \ref{section3} we investigate the unstable modes and the unstable region under the linear perturbation. In section \ref{section4} we present the numerical step in the first subsection \ref{section4.1} and the results in the next subsection \ref{section4.2}.

\section{Einstein-Maxwell-scalar (EMS) theory}\label{section2}

The action of Einstein-Maxwell-scalar theory reads
\begin{equation}
S=\frac{1}{16\pi}\int d^{4}x\sqrt{-g}\left[R-2\nabla_{\mu}\phi\nabla^{\mu}\phi-f(\phi)F_{\mu\nu}F^{\mu\nu}\right].
\label{eq:action}
\end{equation}
where $R$ is the Ricci scalar, $\phi$ the scalar and $F^{\mu\nu}$ the electromagnetic tensor. Here the $f(\phi)$ takes the form $e^{-b\phi^{2}}$ with $b$ a dimensionaless coupling constant, which we consider the cases with $b<0$. The action remains unchanged under the transformation $\phi \rightarrow -\phi$. The equations of motion are obtained by varying the action (\ref{eq:action}) with respect to $g_{\mu\nu},\phi$ and $A_{\mu}$ respectively.
\begin{align}
R_{\mu\nu}-\frac{1}{2}Rg_{\mu\nu}= & 2\left[\partial_{\mu}\phi\partial_{\nu}\phi-\frac{1}{2}g_{\mu\nu}\nabla_{\rho}\phi\nabla^{\rho}\phi+f(\phi)\left(F_{\mu\rho}F_{\nu}^{\ \rho}-\frac{1}{4}g_{\mu\nu}F_{\rho\sigma}F^{\rho\sigma}\right)\right],\\
\nabla_{\mu}\nabla^{\mu}\phi= &-\frac{b}{2}\phi \: e^{-b\phi^{2}}F_{\mu\nu}F^{\mu\nu} \label{eq:equations of scalar},\\
\nabla_{\mu}\left(f(\phi)F^{\mu\nu}\right)= & 0.
\label{eq:equations of electromagnetic}
\end{align}
The right-hand side of eq.\ref{eq:equations of scalar}, which is associated with the coupling function, can be considered as an effective mass term and lead to the spontaneous scalarization of black holes when $b<0$. It is worth pointing out that this model has both RN solution and hairy black hole solution.

\section{Linear perturbation}
\label{section3}

We first study the linear perturbation of the RN black holes in EMS theory and expose three unstable mode branches in the spectrum. The unstable region can also be figured out in the linear level. In the first subsection, we introduce the numerical method for calculating modes and the results are presented in the next subsection.

\subsection{Continued fractions method}
The continued fraction method (CFM) is a powerful approach to solving the eigenvalue problem and is widely adopted to calculate the quasi-normal modes of the black hole perturbation \cite{Leaver,Konoplya2011,Zhang:2015jda} . Its most remarkable advantage is that the accuracy of modes increases with the increase of the order of the Frobenius series. Because RN black hole is one of the solutions of the EMS model, we introduce the perturbation in the background of RN black hole spacetime
\begin{equation}
    ds^{2}= -f(r) dt^{2} + \frac{1}{f(r)} dr^{2} + r^{2} d\Omega^{2},
    \label{eq:RN metric}
\end{equation}
where $f(r)=1-\frac{2M}{r}+\frac{Q^{2}}{r^{2}}$. The scalar perturbation equation of (\ref{eq:equations of scalar}) is given by
\begin{equation}
    \nabla^{\mu}\nabla_{\mu}\delta \phi = -\frac{b}{2}F_{\mu\nu}F^{\mu\nu}\delta \phi=\frac{bQ^2}{r^4}\delta\phi.
    \label{eq:equation of perturbation}
\end{equation}
Note that the effective mass squared $\mu_{\text{eff}}^2=\frac{bQ^2}{r^4}$ is negative when $b<0$. Thus tachyonic instability may be triggered. However, $\mu_{\text{eff}}^2<0$ is not the sufficient condition for the instability. According to the well-known result in quantum mechanics, we should consider the spatial integration of the effective potential of $\delta\phi$ \cite{Buell}. Taking the ansatz $\delta \phi = e^{-i\omega t} \frac{R(r)}{r} Y_{lm}(\theta,\varphi)$ and the redefinition of the tortoise coordinate $dr_{*}=dr/f(r)$, the radial equation (\ref{eq:equation of perturbation}) can be reduced to the the Schr\"odinger-like equations
\begin{eqnarray}
    \frac{d^{2}R(r)}{dr_{*}^{2}}&=&\left[ V(r)-\omega^{2} \right] R(r), \label{eq:Schrodinger-like}\\
    V(r)&=& f(r) (\frac{bQ^{2}}{r^{4}} +\frac{f'(r)}{r}) \label{eq:effective potential},
\end{eqnarray}
where we have fixed $l=0$ in the rest of this paper. The correction of the non-minimal coupling is presented in the term associated with $b$.
We plot the effective potential by the compact coordinate $u=\frac{r-r_{h}}{r}$ in  Fig.\ref{fig:effective potential} where $r_{h}$ is the event horizon radius. The effect of correction is to produce a negative potential well in the effective potential when $-b$ is large enough. The sufficient condition for the instability is then $\int_{-\infty}^{\infty} V dr_\ast = \int_{r_h}^{\infty} \frac{V}{f}dr<0$. For $Q/M=0.8$, the sufficient but not necessary condition for  instability is $b<-5.5$. The well depth increases when $-b$ increases and can accumulate the scalar driving the system away from the scalar-free RN solution.
\begin{figure}[htbp]
    \centering
    \includegraphics[width = 0.5\textwidth]{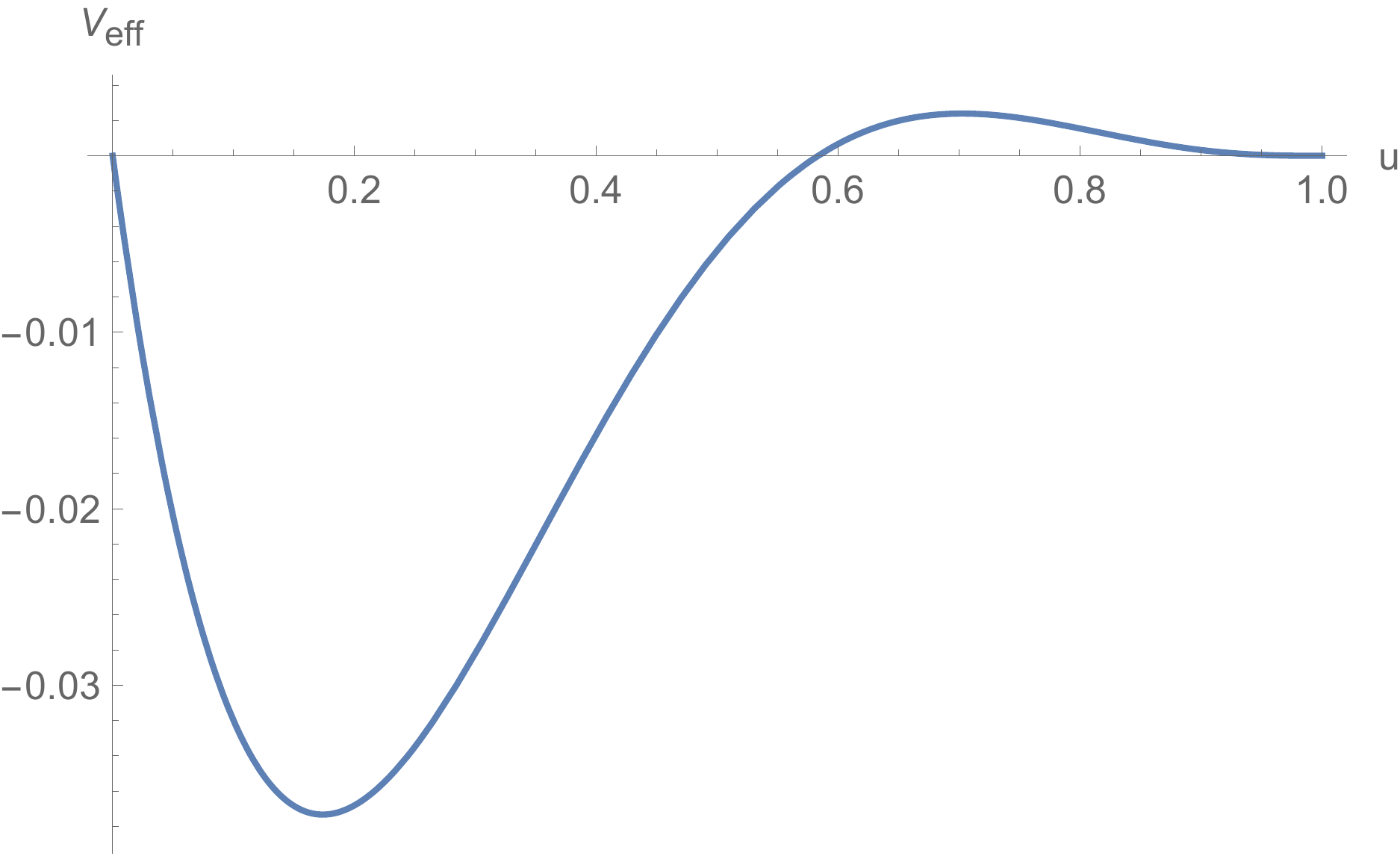}
    \caption{The effective potential with the compact coordinate $u=\frac{r-r_{h}}{r}$. Here we take the parameters $M=1,Q=0.8$.}
    \label{fig:effective potential}
\end{figure}

The asymptotic behavior of equation (\ref{eq:Schrodinger-like}) is
\begin{equation}
    R(r) \sim \left\{
    \begin{array}{ll}
    (r-r_{h})^{-i \frac{r_{h}^{2}\omega}{r_{h}-r_{-}}} ,\ \ & r\rightarrow r_{h} \\
    e^{i\omega r} r^{i\omega} ,\ \ & r\rightarrow \infty
    \end{array}
    \right.
    \label{eq:asymptotic behavior}
\end{equation}
which satisfies purely ingoing boundary condition at the horizon and purely outgoing boundary condition at infinity. Here $r_{-}$ is the radius of the inner horizon of the RN black hole. To implement the CFM, we take the Frobenius series as
\begin{equation}
    R(r)=(r-r_{-})^{i\omega} e^{i\omega r} \left( \frac{r-r_{h}}{r-r_{-}} \right)^{-i \frac{r_{h}^{2}\omega}{r_{h}-r_{-}}}
    \sum_{k=0}^{\infty} a_{k} \left(\frac{r-r_{h}}{r-r_{-}}\right)^{k}.
    \label{eq:Frobenius series}
\end{equation}
Inserting (\ref{eq:Frobenius series}) into (\ref{eq:Schrodinger-like}) where the series is truncated to $N$, we obtain a complicated $N$-term recurrence relation on sequence ($a_{k}$) which can be reduced to the $3$-term recurrence relation numerically.
\begin{eqnarray}
    c^{(3)}_{0,i}a_{i}+c_{1,i}^{(3)}a_{i-1}+c_{2,i}^{(3)}a_{i-2}&=&0, \quad \textrm{for} \; i>1, \nonumber \\
    c^{(3)}_{0,1}a_{1}+c_{1,1}^{(3)}a_{0} &=& 0.
    \label{eq:3 term recurrence relation}
\end{eqnarray}
The coefficients of (\ref{eq:3 term recurrence relation}) can be used to construct the continue fractions
\begin{equation}
        g(\omega)=c^{(3)}_{1,1}-\frac{c^{(3)}_{0,1} c^{(3)}_{2,2}}{c^{(3)}_{1,2}- \frac{c^{(3)}_{0,2}c^{(3)}_{2,3}}{c^{(3)}_{1,3}- \cdots }}.
        \label{eq:continued fraction}
\end{equation}
Fixing the parameters ($Q,b$), $g(\omega)$ is zero if $\omega$ takes the values of the quasi-normal frequencies. The eigenvalue problem is immediately turned into a problem of function $g(\omega)$ searching for zeros on the complex $\omega$ plane.

\subsection{Modes}
\label{subsection3.2}

\begin{figure}[htbp]
    \centering
    \includegraphics[width = 0.32\textwidth]{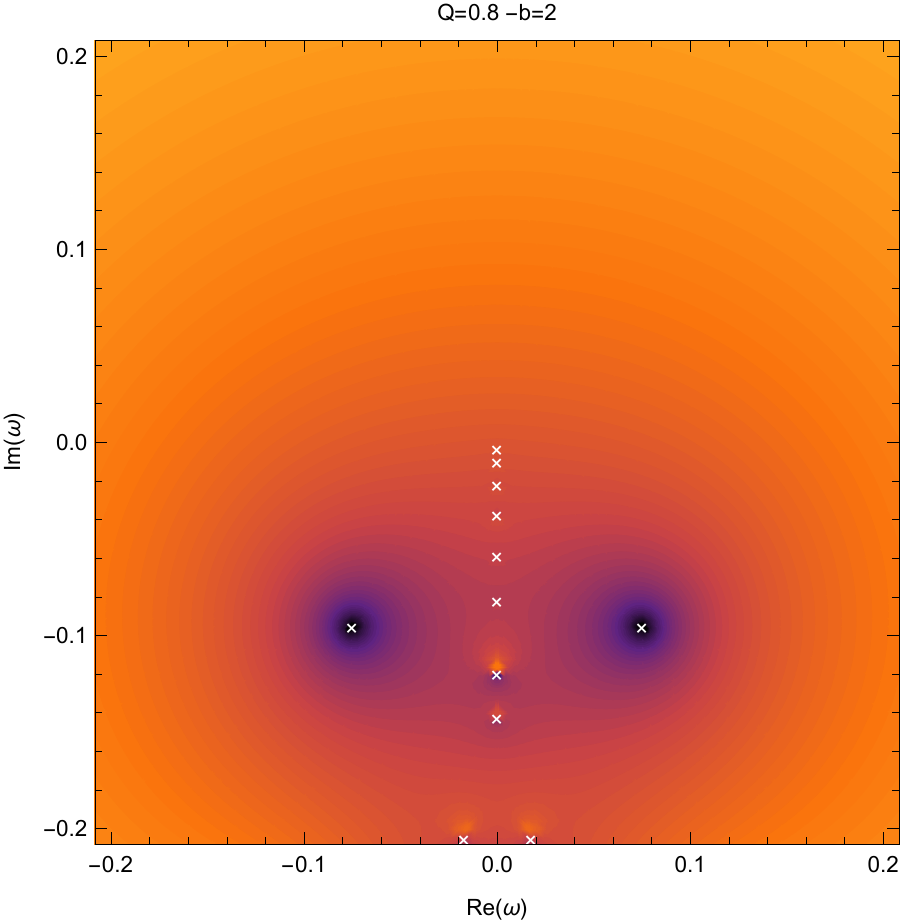}
    \includegraphics[width = 0.32\textwidth]{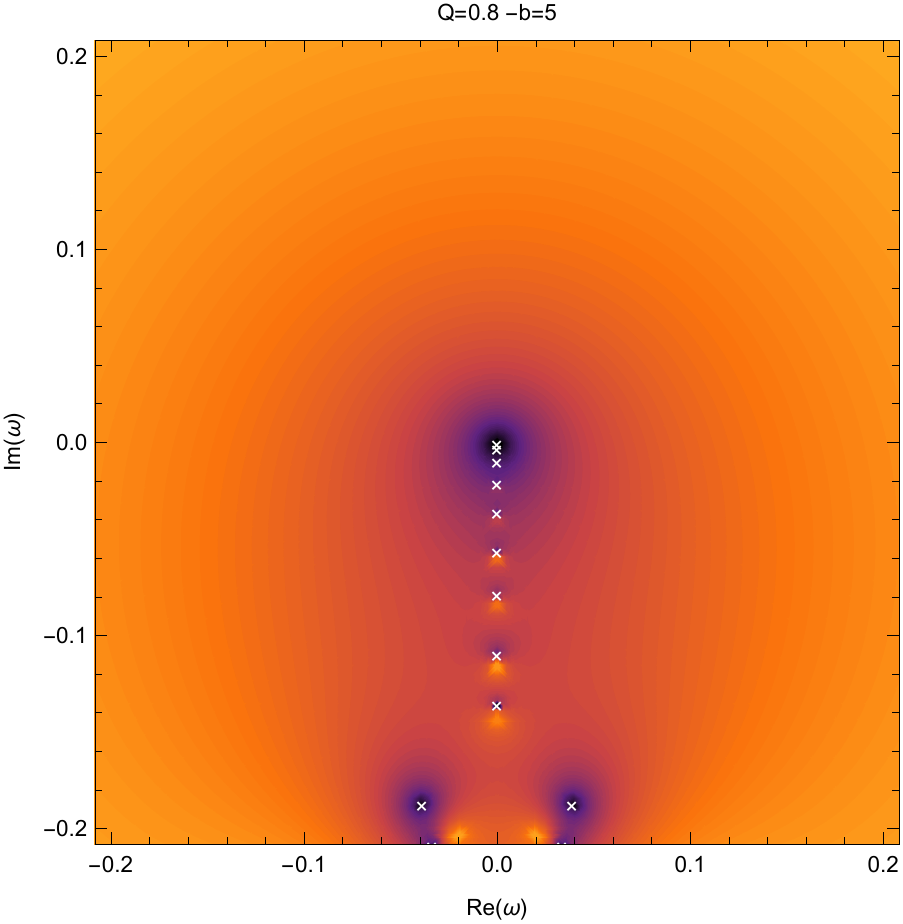}
    \includegraphics[width = 0.32\textwidth]{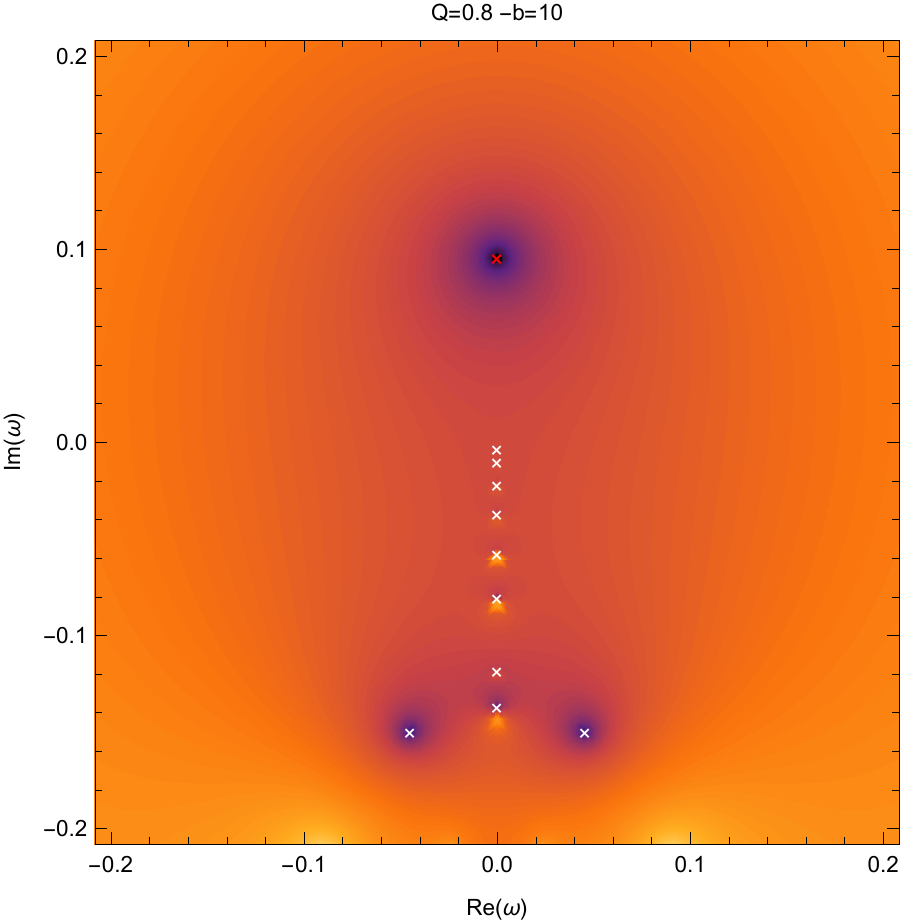}
    \caption{Continued fraction values of the RN black hole in EMS theory when $M=1,Q=0.8$. Left, middle and right panels for $-b=2,5,10$, respectively. 
 Quasinormal modes locate at the crosses.}
    \label{fig:CFM}
\end{figure}

We draw the continued fraction values of $\log_{10}|g(\omega)|$ with respect to $b$ on the complex plane when $Q/M=0.8$ in Fig.\ref{fig:CFM}. For $-b=2$, all the QNMs of the RN black hole in EMS theory have negative $\omega_I$, i.e., there are no unstable modes. The first unstable mode appears when $-b=5$. As $-b$ increases further, more unstable modes appear. This is consistent with the analysis of effective potential. The second unstable mode appears when $-b=26$. All the unstable modes locate at the image axis, that is, they are pure imaginary modes.

\begin{figure}[htbp]
    \centering
    \includegraphics[width = 0.45\textwidth]{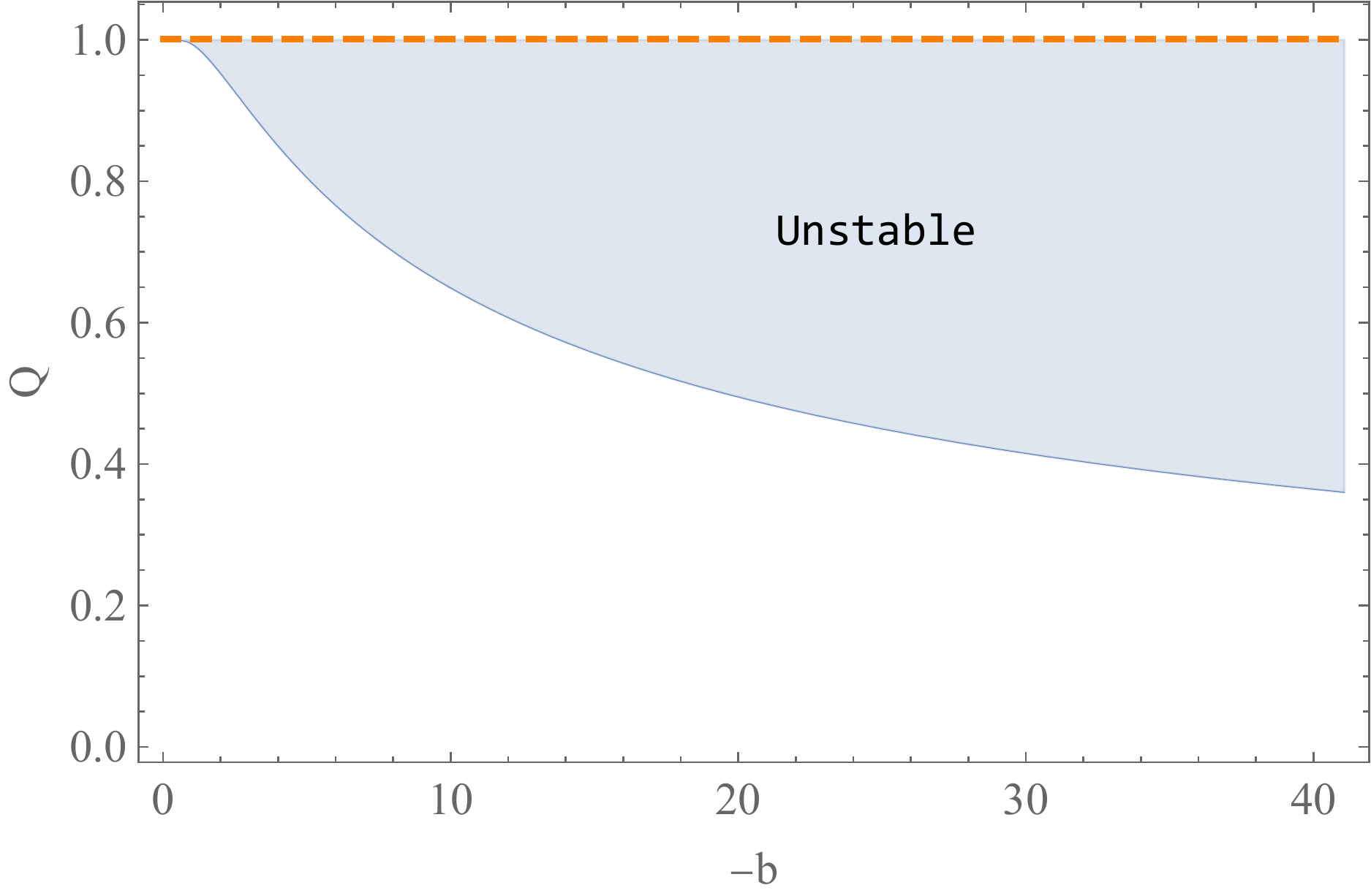}
    \includegraphics[width = 0.45\textwidth]{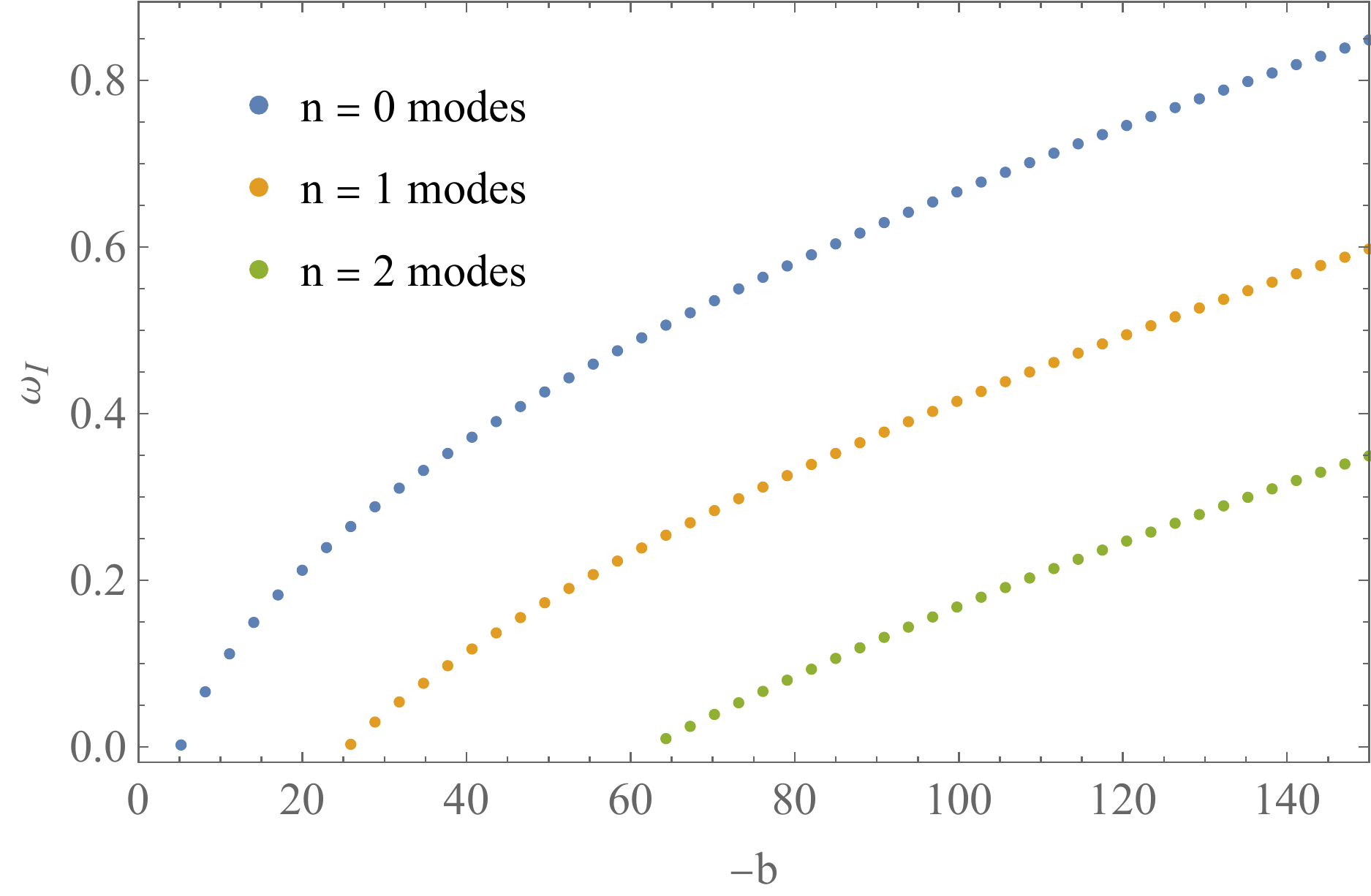}
    \caption{The left panel is the unstable region of parameter plane for $Q,-b$. The right panel is the imaginary parts of unstable modes with different $-b$ when fixing $Q=0.8$.}
    \label{fig:unstable region and migration modes}
\end{figure}

The parameter region where RN black holes have unstable modes is shown in the left panel Fig.\ref{fig:unstable region and migration modes}. The unstable region obtained by linear perturbation is in good agreement with the similar figure presented in \cite{Herdeiro:2018wub}. However, we can not gain the unstable region beyond the critical charge $Q_{c}=1$ because there is no definition of the asymptotic behavior for a bare singularity.

The imaginary parts of the unstable modes with different $-b$ are shown in the right panel of Fig.\ref{fig:unstable region and migration modes} where $Q=0.8$. The first branch of unstable modes first appears while beyond the critical $-b=5$. With increasing of $-b$, the second branch emerges at about $-b=26$ and the third branch emerges at about $-b=63$. These higher overtone modes are associated with the different bound states of the perturbation equation \cite{Bosch:2019anc}.

\section{Nonlinear evolution}
\label{section4}

We simulate the nonlinear evolution of EMS black hole in spherically symmetric spacetime by taking the Painlev\'e-Gullstrand(PG)-like coordinates ansatz
\begin{equation}
    ds^{2} =-\left(1-\zeta^{2}\right)\alpha^{2}dt^{2}+2\zeta\alpha dtdr+dr^{2}+r^{2}(d\theta^{2}+\sin^{2}\theta d\phi^{2}).
    \label{eq:metric ansatz}
\end{equation}
There are $\alpha=1,\zeta=\sqrt{\frac{2M}{r}-\frac{Q^{2}}{r^{2}}}$ for RN black hole. The PG coordinates haven been used to study the black hole dynamics  numerically \cite{Ripley2020,Ripley2019,Zhang:2021ybj,Zhang2021b}. It is regular on the horizon and thus appropriate for the long-term simulation of the black hole dynamics. We take the gauge potential
\begin{equation}
A_{\mu}dx^{\mu}=A(t,r)dt,
\end{equation}
and introduce following auxiliary variables
\begin{equation}
\Phi=\partial_{r}\phi,\ \ \ P=\frac{1}{\alpha}\partial_{t}\phi-\zeta\Phi,\ \ \ E=\frac{1}{\alpha}\partial_{r}A.
\label{eq:Phi,P,E}
\end{equation}
From the Maxwell equations (\ref{eq:equations of electromagnetic}), we have
\begin{equation}
\partial_{r}\left(r^{2}f(\phi)E\right)=0,\ \ \ \partial_{t}\left(r^{2}f(\phi)E\right)=0.
\end{equation}
Then we have
\begin{equation}
E=\frac{Q}{r^{2}f(\phi)},
\end{equation}
in which $Q$ is a constant interpreted as the electric charge. The Einstein equations become
\begin{align}
\partial_r\alpha= & -\frac{rP\Phi\alpha}{\zeta}, \label{eq:alpha}\\
\partial_r\zeta= & \frac{r}{2\zeta}\left(\Phi^{2}+P^{2}+\Lambda\right)+\frac{Q^{2}}{2r^{3}\zeta f(\phi)}+rP\Phi-\frac{\zeta}{2r},\label{eq:zetadr}\\
\partial_{t}\zeta= &\frac{r\alpha}{\zeta}\left(P+\Phi\zeta\right)\left(P\zeta+\Phi\right).
\label{eq:zetat}
\end{align}
The scalar field equation is given by
\begin{align}
\partial_{t}\phi & =\alpha\left(P+\Phi\zeta\right),\label{eq:phit}\\
\partial_{t}P  & =\frac{\left(\left(P\zeta+\Phi\right)\alpha r^{2}\right)'}{r^{2}}+\frac{\alpha}{2}\frac{f'(\phi)Q^{2}}{r^{4}f^{2}(\phi)}.\label{eq:Pt}
\end{align}

We solve the system of equations(\ref{eq:alpha}-\ref{eq:Pt}) in the frame of fully nonlinear evolution. The numerical steps are shown in subsection \ref{section4.1} and the results are presented in subsection \ref{section4.2}. In the rest of this paper, we fix $M=1$ and $Q=0.8$ except specifically mentioned. The Misner-Sharp mass is
\begin{equation}
M_{MS}(t,r)=\frac{r}{2}\left(1-g^{\mu\nu}\partial_{\mu}r\partial_{\nu}r\right)=\frac{r}{2}\zeta(t,r)^{2},
\end{equation}
which tends to be the spacetime mass when $r\to\infty$. We also study the black hole irreducible mass $M_{h}=\sqrt{\frac{A}{4\pi}}$ in which $A$ is the area of the apparent horizon. 

\subsection{Numerical step}
\label{section4.1}

Here we first discuss the boundary conditions for the nonlinear equations of evolution. The auxiliary freedom in the metric ansatz (\ref{eq:metric ansatz}) allows us to
fix
\begin{equation}
    \alpha |_{r\rightarrow \infty} = 1
\end{equation}
by rescaling the time coordinate. The other metric function $\zeta$ tends to $\sqrt{\frac{2M}{r}}$ when $r \rightarrow \infty$. We take the initial profile for the scalar field as
\begin{equation}
    \phi = \kappa e^{-(\frac{r-6 r_{h}}{r_{h}})^{2}} ,\ \  P=0,
\end{equation}
where $r_{h}$ is the horizon of the initial black hole solution and $\kappa$ is of order $10^{-7}$ such that we can neglect the back reaction of the scalar field to the initial background spacetime. The $\Phi$ is given by (\ref{eq:Phi,P,E}). Plugging the initial $\phi, \Phi, P$ into the constraint equation (\ref{eq:zetadr}), the initial $\zeta$ can be obtained by the Newton-Raphson method. Then the initial $\alpha$ can be obtained from (\ref{eq:alpha}).

Given parameters ($M, Q, b$), the boundary conditions and initial profiles ($\phi, \Phi, P, \zeta, \alpha$), the numerical simulation process is following: we get $\zeta,\phi,P$ on the next time slice from (\ref{eq:zetat},\ref{eq:phit},\ref{eq:Pt}); then we can get $\Phi,\alpha$ from (\ref{eq:Phi,P,E},\ref{eq:alpha}). We iterate these procedure to gain all the following time slices.

We work out in the radial computational region ranges from $r_{0}$ to $\infty$, where $r_{0}=0.7 r_{h}$. The $r_{0}$ always lies in the apparent horizon, which is guaranteed by the initial apparent horizon $r_{h}$. In principle, the information would not affect the region outside the horizon. The computational region is compactified by a coordinate transformation $z=\frac{r}{r+M}$ which ranges at $(z_{0},1)$. We use the finite difference method in the radial direction and discrete the radial direction by a uniform grid with $2^{11}\sim 2^{12}$ points. We evolve the system through the fourth-order Runge-Kutta method. We also use the Kreiss-Oliger dissipation to stabilize the numerical evolution.

\begin{figure}[htbp]
    \centering
    \includegraphics[width = 0.4\textwidth]{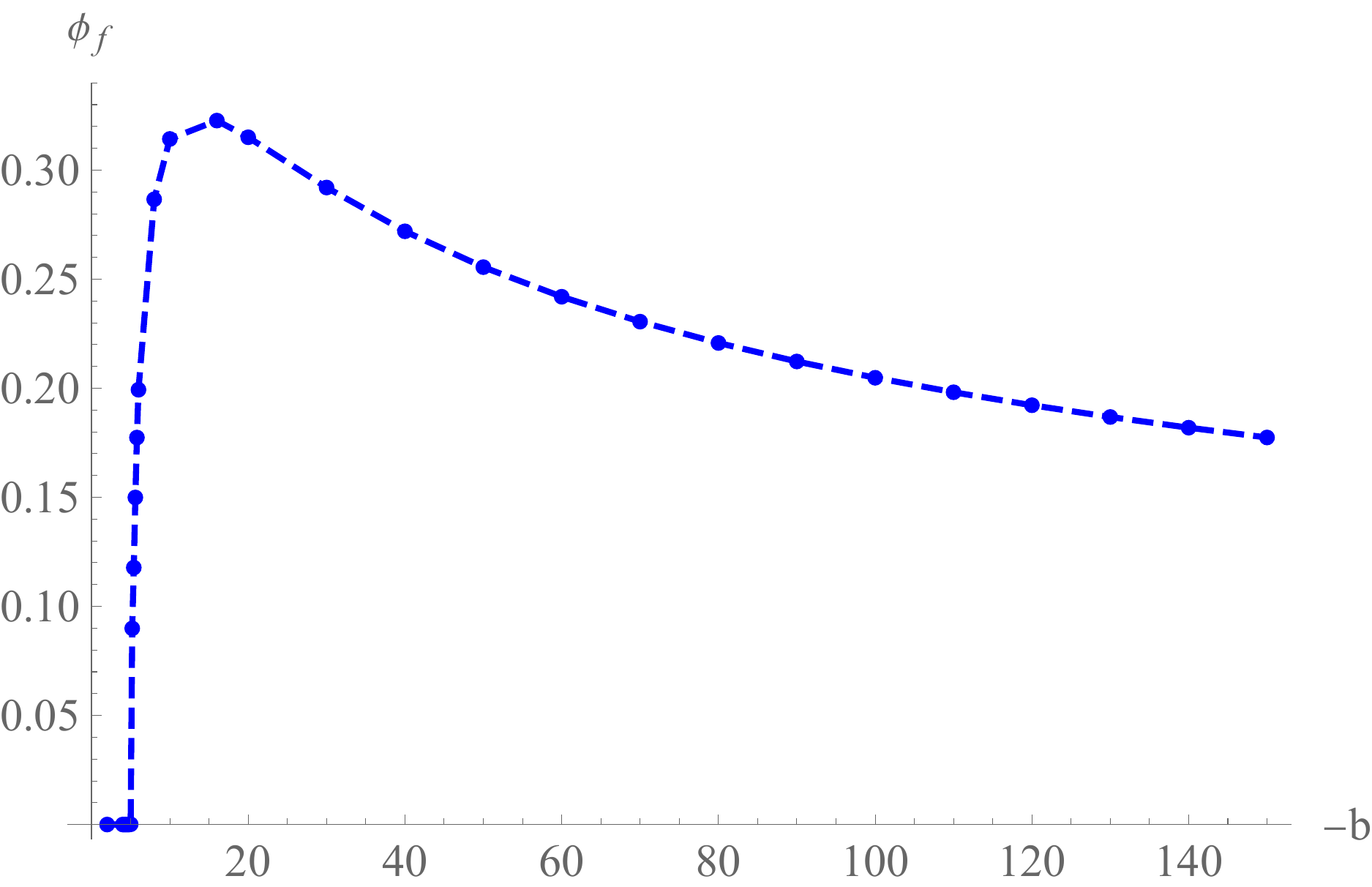}
    \includegraphics[width = 0.4\textwidth]{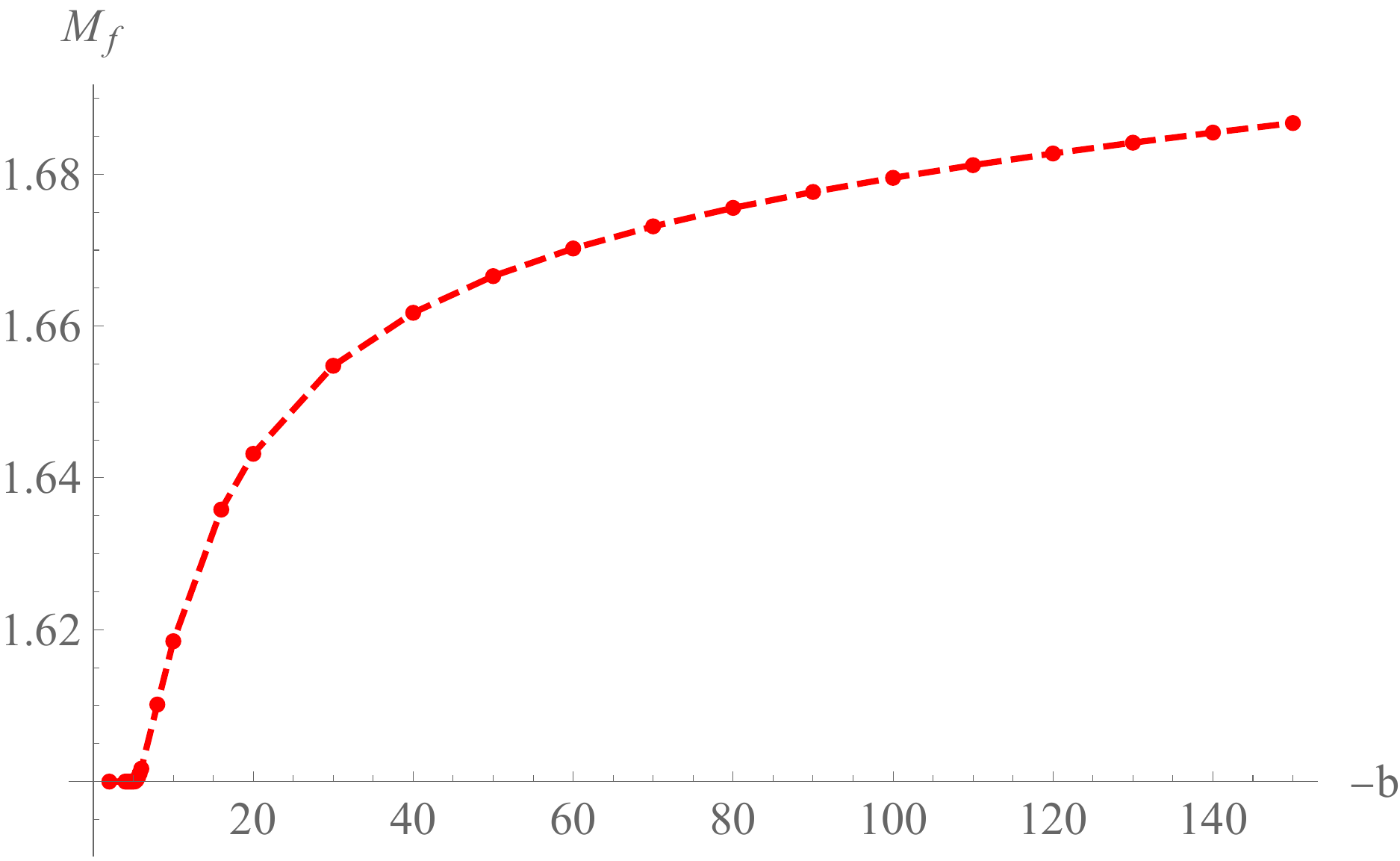}
    \caption{The final value of scalar field (left) and the irreducible mass (right) at horizon.}
    \label{fig:phif Mf vs b}
\end{figure}

\subsection{Results}
\label{section4.2}

We first investigate the final value $\phi_{f}$ of the scalar field and the black hole irreducible mass $M_{f}$ with respect to $-b$. The results are shown in Fig.\ref{fig:phif Mf vs b}. The scalar is inhibited while $-b$ is below the critical value $-b=5$, and we can deduce that this system evolves into the RN black hole. A similar situation occurs in the final irreducible mass $M_{f}$. When $-b$ exceeds the critical value, $\phi_{f}$ increases rapidly and then decreases with $-b$. This shows that the scalar field absorbs electromagnetic energy through the coupling term $e^{-b\phi^{2}}F_{\mu\nu}F^{\mu\nu}$. The $M_{f}$ increases monotonically with $-b$ increasing, which means that more energy of the scalar field is absorbed by the black hole for stronger coupling.

\begin{figure}[htbp]
    \centering
    \includegraphics[width = 0.4\textwidth]{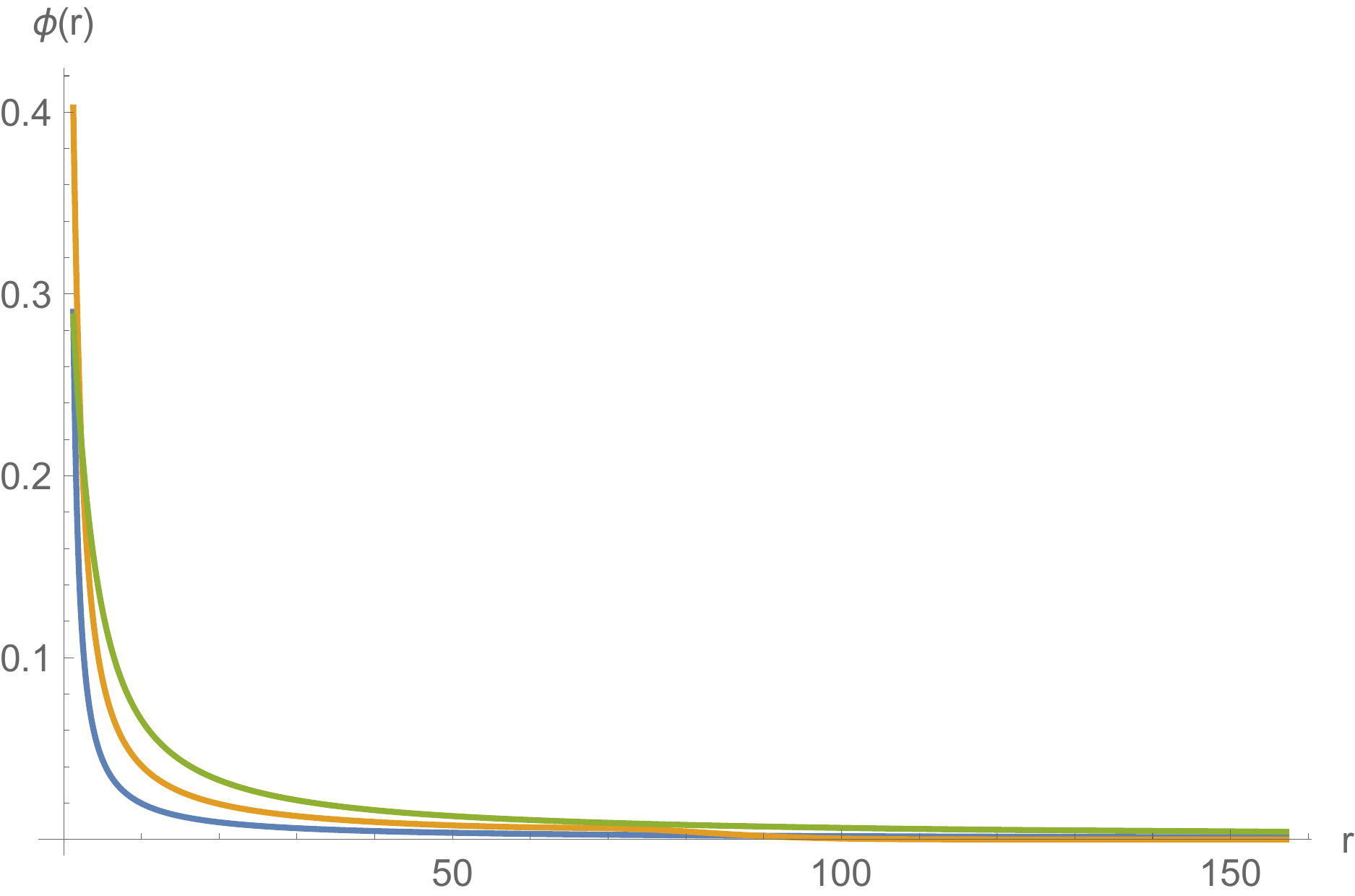}
    \includegraphics[width = 0.4\textwidth]{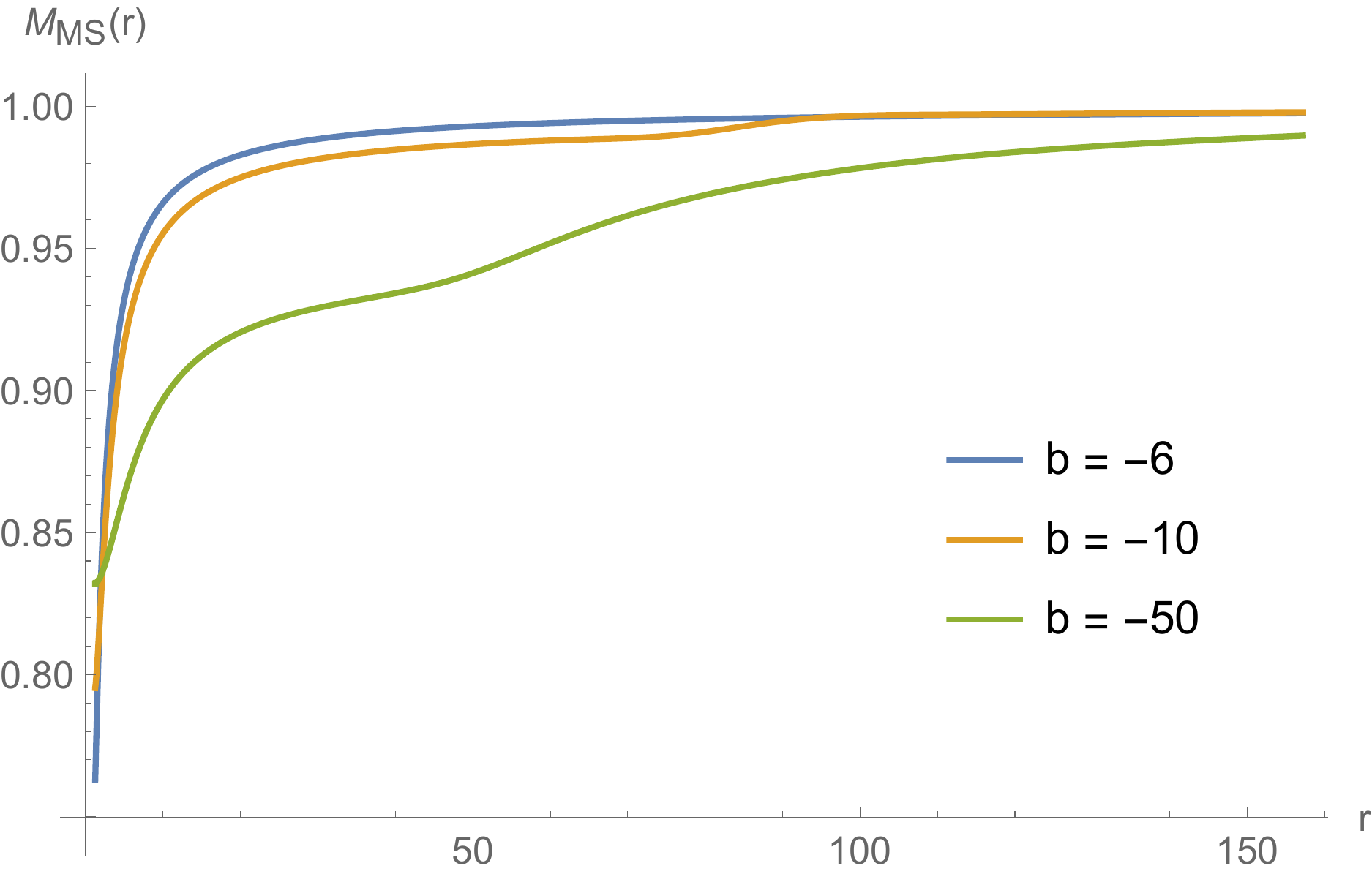}
    \caption{The left and right plots are the final profile of the scalar field and the Misner-Sharp mass with different $b$.}
    \label{fig:phi MMS vs r}
\end{figure}

In Fig.\ref{fig:phi MMS vs r} we show the radial profiles of the scalar field $\phi$ and the Misner-Sharp mass $M_{MS}$ in the final state. The scalar field condenses near the horizon which is more obvious with small $-b$ and the Misner-Sharp mass distribution confirms this. The coupling term $-\frac{b}{2}\phi \: e^{-b\phi^{2}}F_{\mu\nu}F^{\mu\nu}$ of equation (\ref{eq:equations of scalar}) hence generates an effective repulsive potential. This potential decreases rapidly with $r$ in the form of $e^{-b\phi^{2}}$ and drives away the scalar field from the black hole. This effect is enhanced by larger $-b$, which can be used as an explanation for the decay with large $-b$ of the final value of $\phi_{h}$ in the left panel of Fig.\ref{fig:phif Mf vs b}.

\begin{figure}[htbp]
    \centering
    \includegraphics[width = 0.4\textwidth]{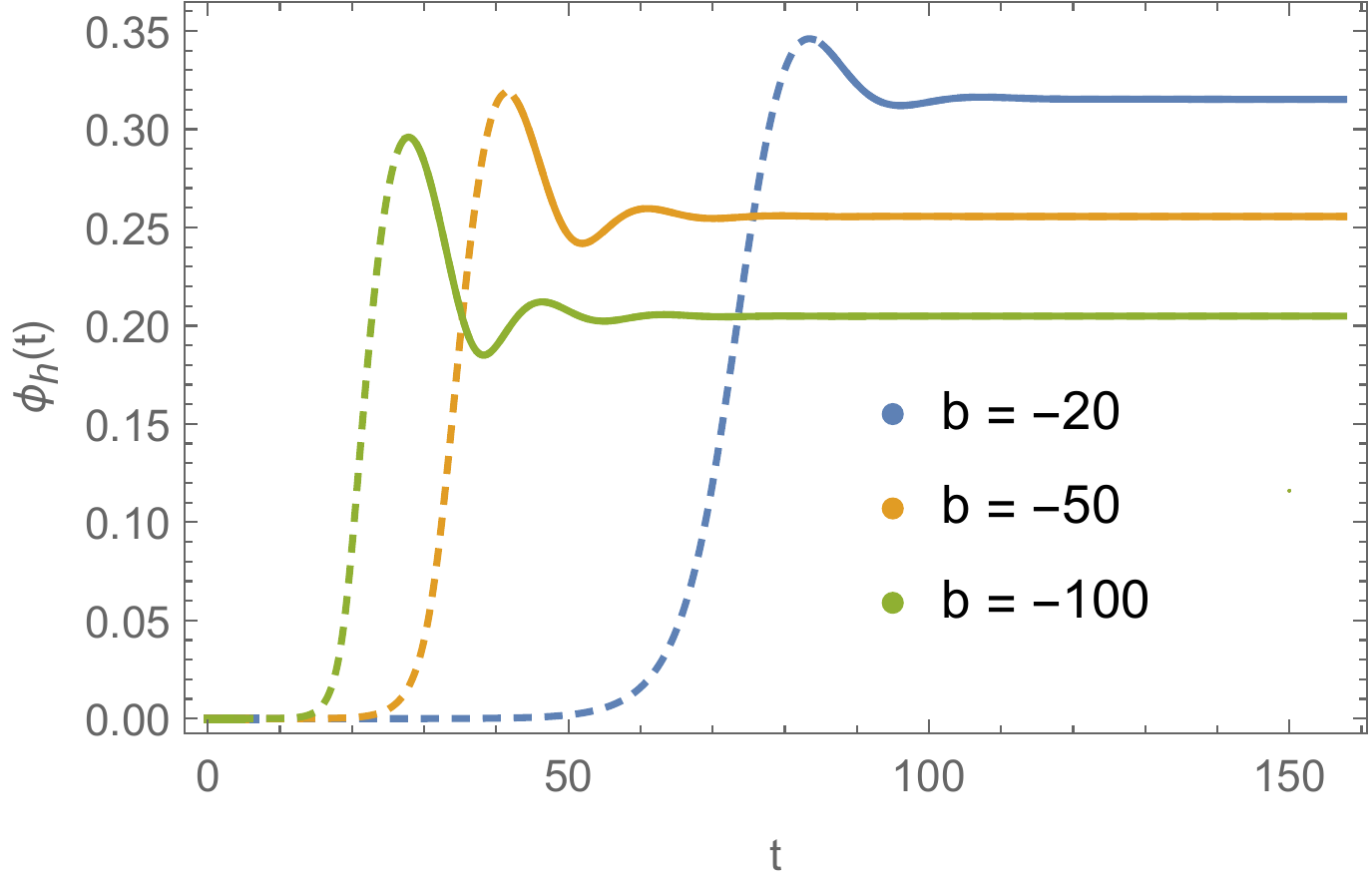}
    \includegraphics[width = 0.4\textwidth]{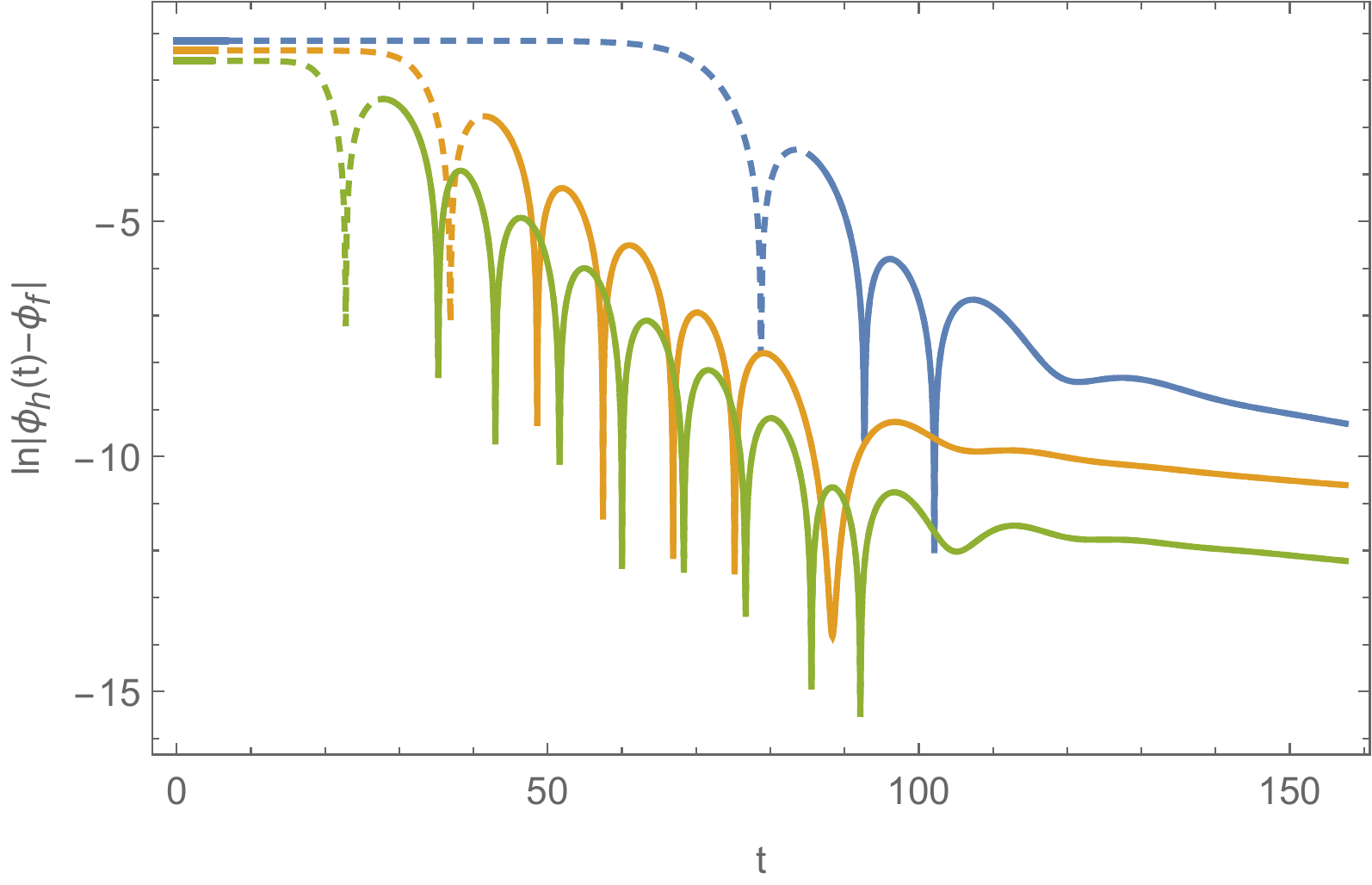}
    \includegraphics[width = 0.4\textwidth]{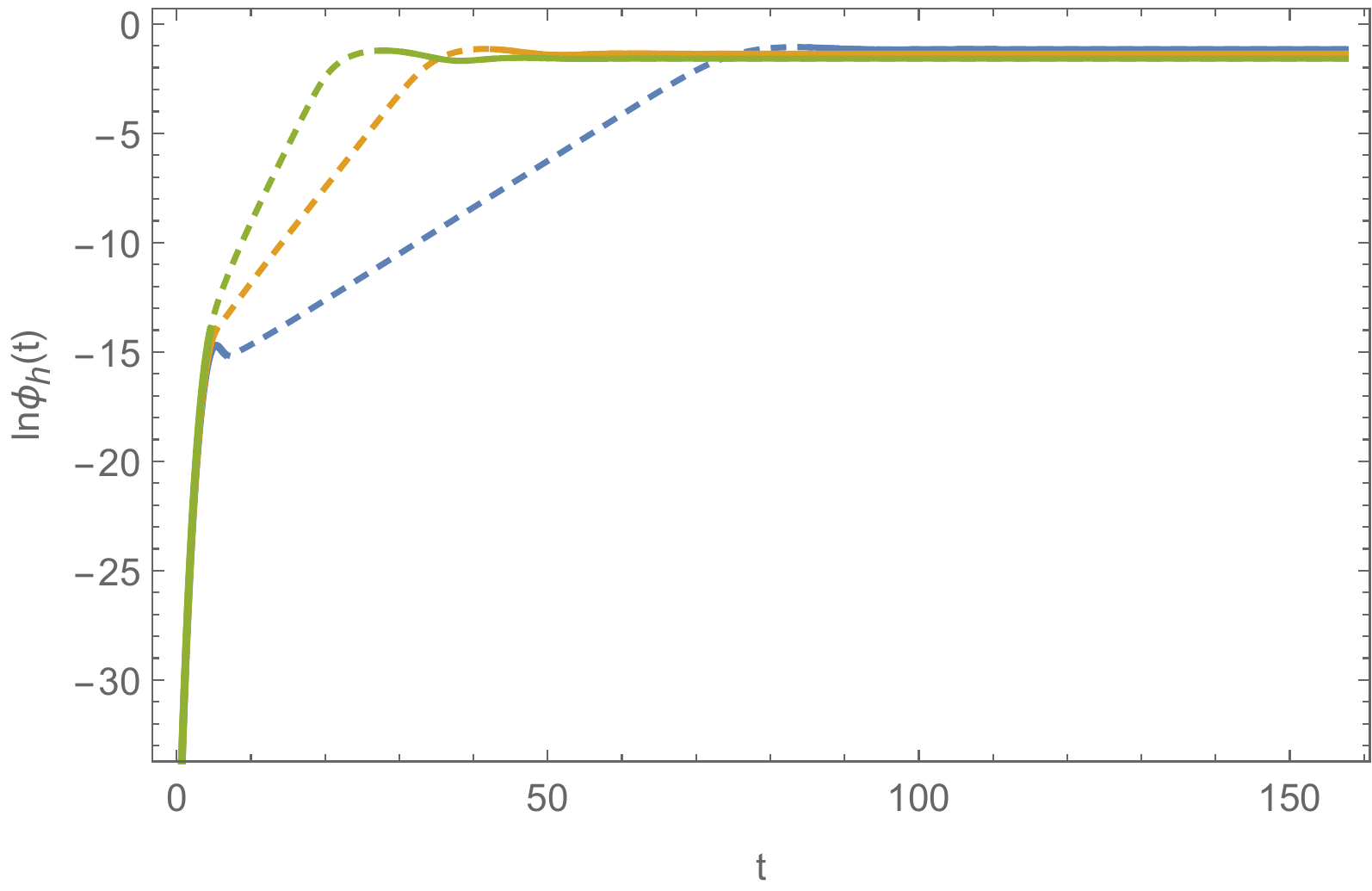}
    \includegraphics[width = 0.4\textwidth]{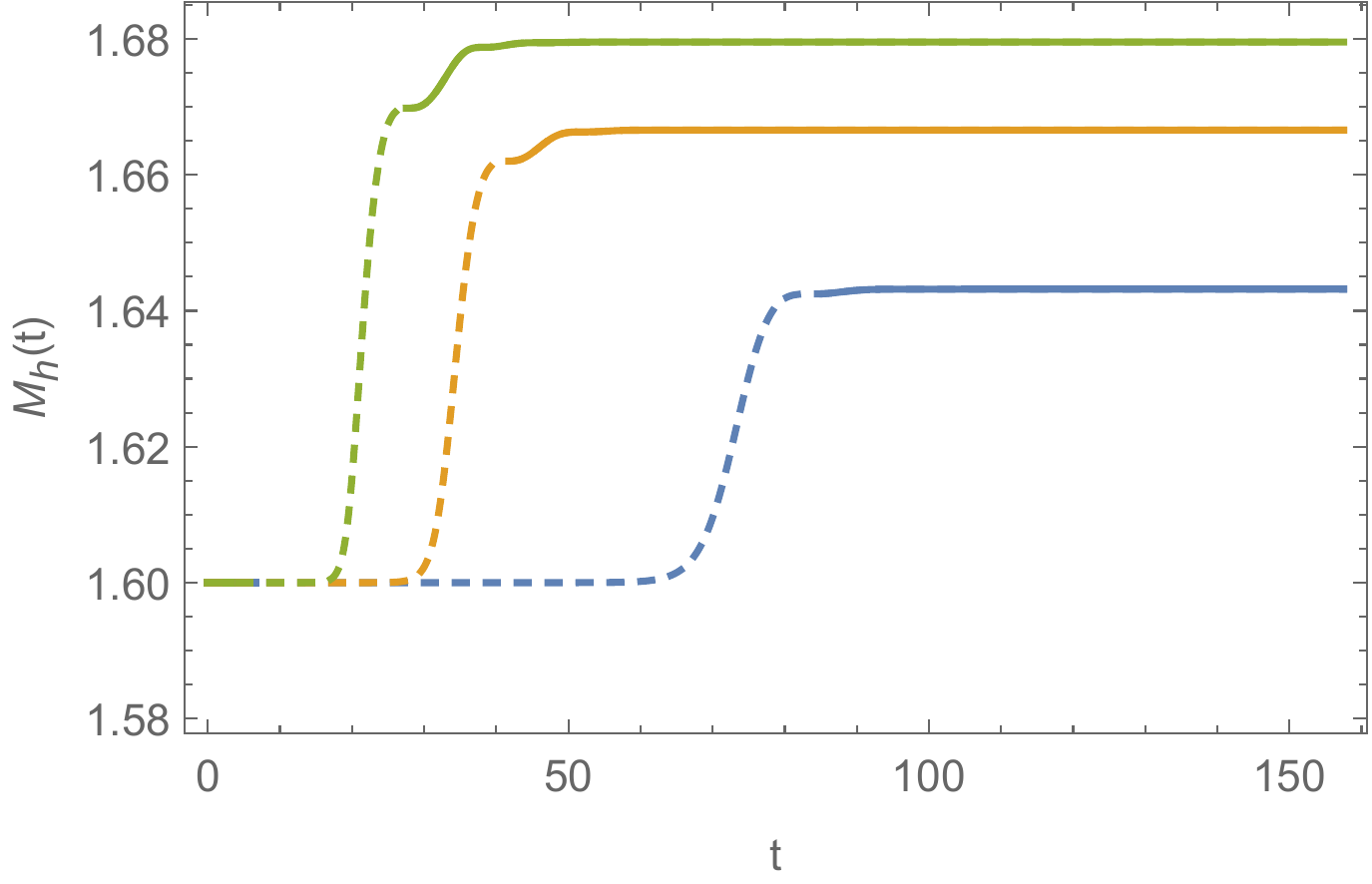}
    \caption{The upper left plot is the evolution of the scalar field at the horizon; the upper right plot is the logarithm of $|\phi_{h}-\phi_{f}|$ vs $t$; the lower left plot is the logarithm of scalar vs $t$; the lower right plot is the irreducible mass at the horizon with time evolution. All plots share the same time interval and the three line markers represent different evolutionary stages.}
    \label{fig:phih Mh vs t}
\end{figure}

The evolution of the scalar field $\phi_{h}$ on the apparent horizon and the irreducible mass $M_{h}$ share the same characteristics as shown in Fig.\ref{fig:phih Mh vs t}. We take multiple perspectives to observe the evolution of the scalar field and effectively distinguish each evolution stage by comparing it with linear perturbation. There are three evolutionary stages: wave packet transmission stage, exponential growth stage, and oscillation convergence stage, which are signed by a thick line, dashed line, and thin line respectively.

\begin{figure}[htbp]
    \centering
    \includegraphics[width = 0.435\textwidth]{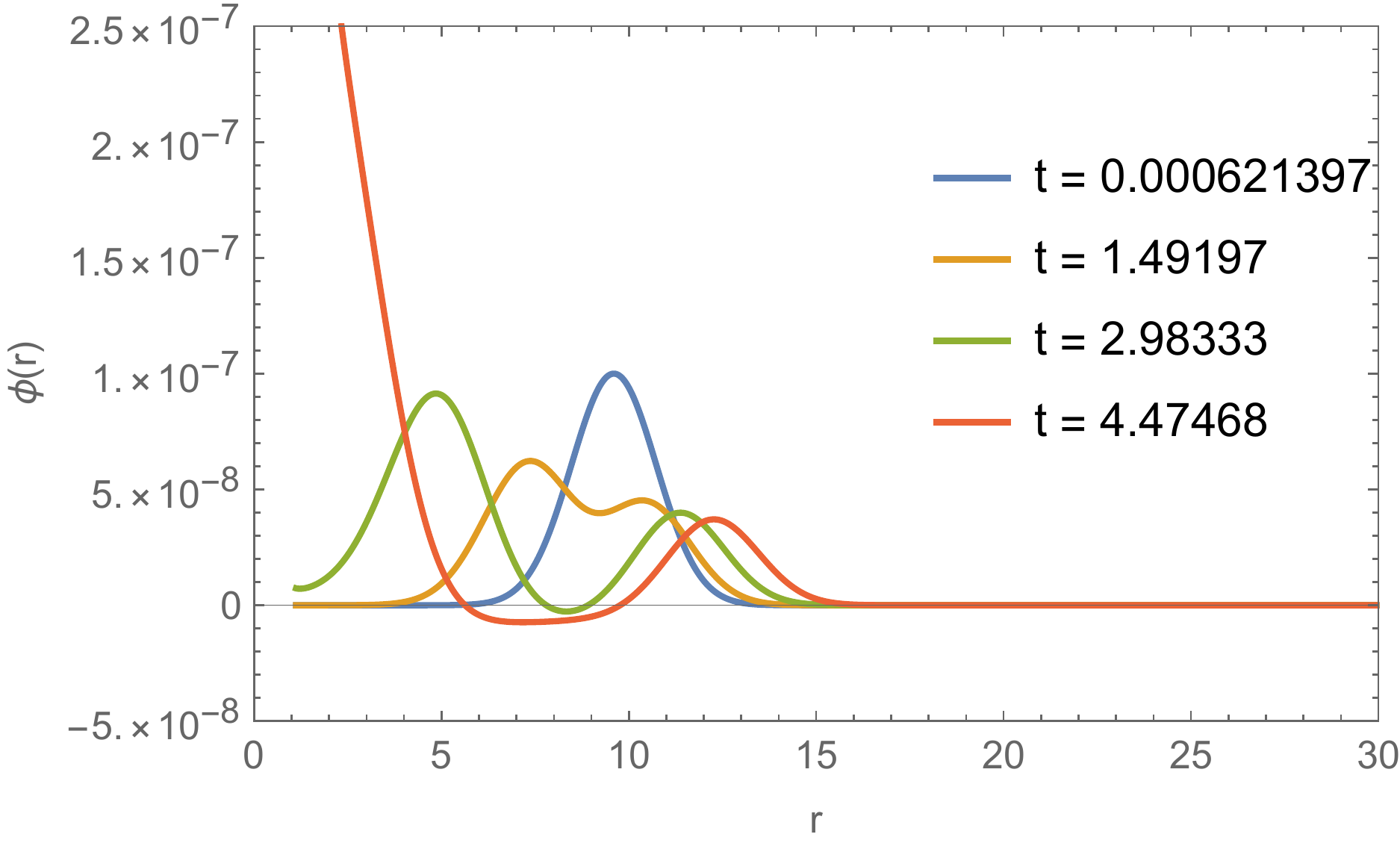}
    \includegraphics[width = 0.4\textwidth]{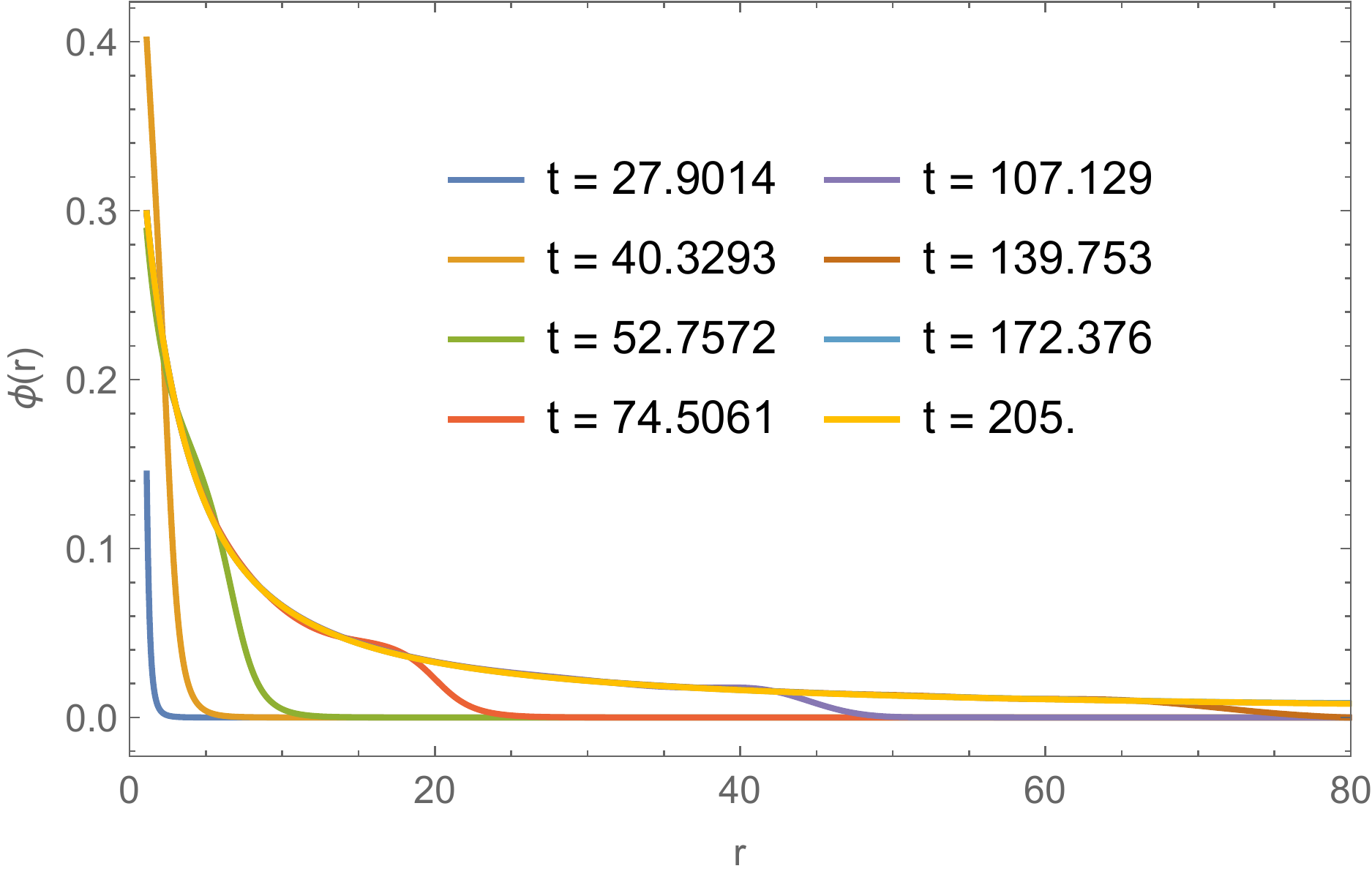}
    \caption{The profile of $\phi(r)$ at the early time (left) and the remaining time (right) where we fix $-b=50$.}
    \label{fig:phi r vs t}
\end{figure}

\begin{figure}[htbp]
    \centering
    \includegraphics[width = 0.6\textwidth]{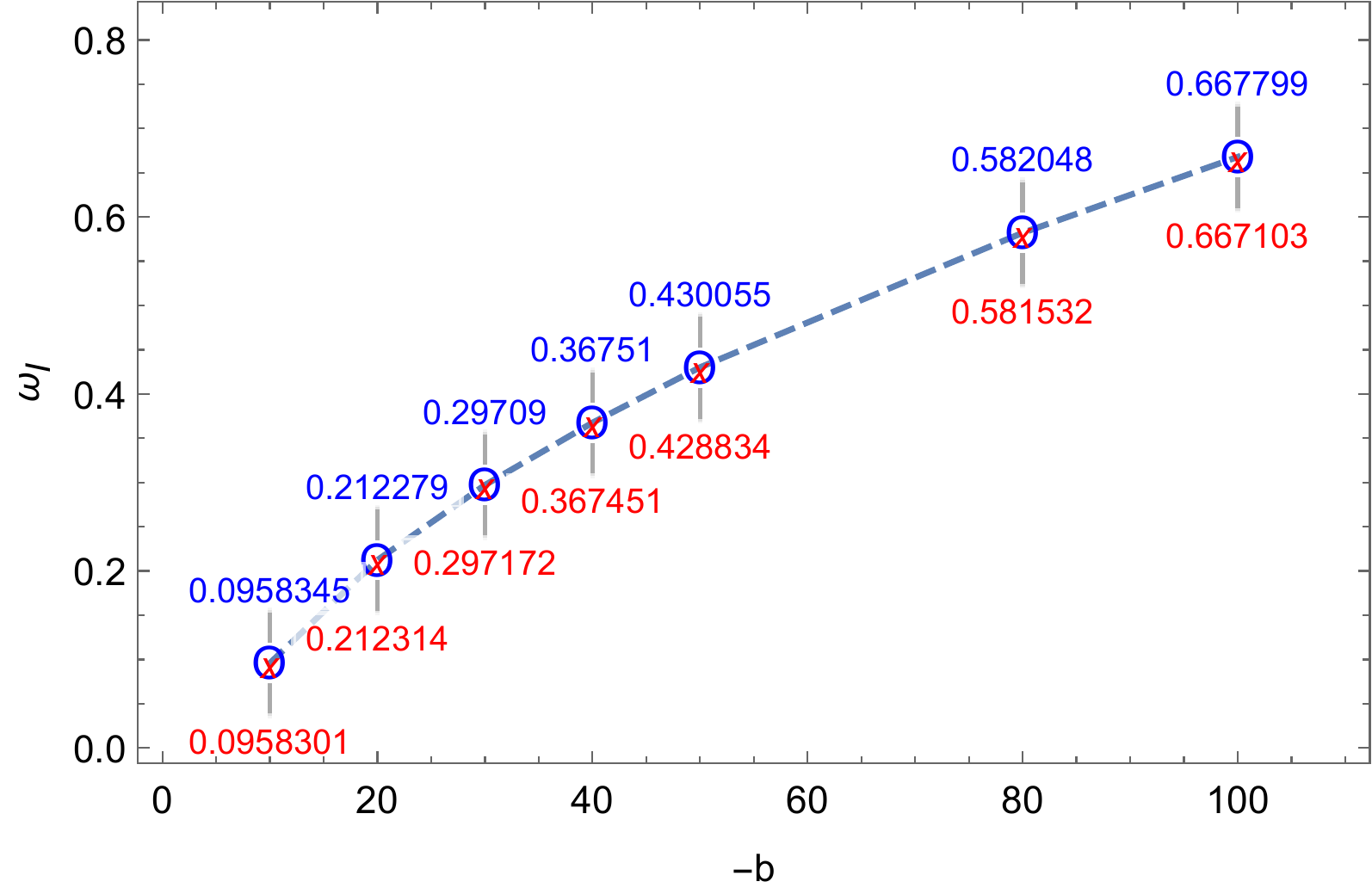}
    \caption{The comparison between the fitting slope (blue) for nonlinear evolution and the fundamental modes (red) for linear perturbation with different $-b$.}
    \label{fig:fitmodes and modes vs b}
\end{figure}

At early time the evolution values ($\phi_{h},M_{h}$) are dominated by the propagation of the initial wave packet, which corresponds to the early evolution of Fig.\ref{fig:phih Mh vs t}. The clearer figure is displayed in the left pane of Fig.\ref{fig:phi r vs t}, which is the radial waveform at the early time. The initial wave packet is separated into two parts: the ingoing wave approaches the event horizon and increases with time; the outgoing wave dissipates to infinity. When the wave packet has reached the horizon, the scalar field increases exponentially, which corresponds to the dashed line segments in Fig.\ref{fig:phih Mh vs t}. The irreducible mass of a black hole also increases exponentially. The scalar evolution in this stage can be fitted well by the exponential function $e^{\omega_I t}$. The comparison between $\omega_I$ from nonlinear evolution and the fundamental modes from the linear perturbation is shown in Fig.\ref{fig:fitmodes and modes vs b} with different $-b$. This excellent coincidence represents that the exponential growth at this stage is dominated by the unstable mode for linear perturbation.

The third stage begins after the highest point of each line in the upper left panel of Fig.\ref{fig:phih Mh vs t}. It corresponds to a decaying oscillation period, which is clear in the upper left and upper right panels of Fig.\ref{fig:phih Mh vs t}. In the right panel of Fig.\ref{fig:phi r vs t}, when the scalar reaches its maximum (the brown line at time $t=40.3293$), the scalar field gathers in a very narrow range near the event horizon. After that, the value of the scalar falls back and spread to infinity. Part of the energy of the scalar field is absorbed by the black hole, which increases the black hole area again as in the lower right panel of Fig.\ref{fig:phih Mh vs t}.

\section{Summaries}
We analyzed the linear scalar perturbation of the RN black hole in EMS theory with nonlinear coupling function $e^{-b\phi^2}$. The nonlinear coupling introduces a negative potential well for the scalar outside the horizon for negative $b$. When the radial integration of the effective potential of the scalar is negative, the tachyonic instability can be triggered. It drives the system away from the scalar-free RN solution, leads to a scalarized charged black hole solution finally. Most of the scalar is accumulated in the potential well. The potential well increases with $-b$. Thus the black hole is dressed heavier for larger $-b$. We analyzed the QNMs structure of the scalar perturbation using the continued fraction method. All the unstable modes have a purely imaginary part. For large $-b$, the overtones can also be unstable. We get the unstable parameter region of the RN black hole in EMS theory. It is in good agreement with that obtained in \cite{Herdeiro:2018wub}.

We also studied the fully nonlinear evolution of the unstable RN black hole after a small perturbation in EMS theory. The evolution endpoint of scalar field $\phi_{f}$ and the irreducible mass $M_{f}$ with different $b$ are presented. We show the radial distribution of $\phi(r)$ and $M_{\textrm{MS}}(r)$ of the final state. Through four different perspectives, we conclude that the scalar evolution can be divided into three stages: wave packet transmission stage, exponential growth stage, and oscillation convergence stage. The exponential coefficient of the second stage coincides well with the unstable mode from the linear analysis. We find that the fundamental unstable mode dominates this stage and the overtones do not. The third stage can be viewed as the perturbation of the final scalarized black hole.

\section*{Acknowledgments }

Peng Liu would like to thank Yun-Ha Zha for her kind encouragement during this work. This research is supported by the National Key R\&D Program of China under Grant No.2020YFC2201400, the Natural Science Foundation of China under Grant Nos.11805083, 11905083, 12005077, and Guangdong Basic and Applied Basic Research Foundation under Grant No.2021A1515012374.

\end{document}